\documentclass[prl, preprint]{revtex4-1}

\usepackage{graphicx}
\usepackage{dcolumn}
\usepackage{bm}
\usepackage[dvipsnames]{xcolor}
\usepackage{amsmath}
\DeclareMathOperator{\sinc}{sinc}

\begin{document}


\title{Mean-field theory of the uniaxial ferroelectric smectic A liquid crystal phase} 

\author{Aitor Erkoreka}
\affiliation{Department of Physics, Faculty of Science and Technology, University of the Basque Country UPV/EHU, Bilbao, Spain}

\author{Mingjun Huang}
\author{Satoshi Aya}
\affiliation{South China Advanced Institute for Soft Matter Science and Technology (AISMST), School of Molecular Science and Engineering, South China University of Technology, Guangzhou, China}
\affiliation{Guangdong Provincial Key Laboratory of Functional and Intelligent Hybrid Materials and Devices, South China University of Technology, Guangzhou, China}

\author{Josu Martinez-Perdiguero}
\email{jesus.martinez@ehu.eus}
\affiliation{Department of Physics, Faculty of Science and Technology, University of the Basque Country UPV/EHU, Bilbao, Spain}

\date{\today}

\begin{abstract}
Following the groundbreaking discovery of the ferroelectric nematic liquid crystal phase (N$_{\text{F}}$), a series of closely-related new polar phases have also been found. An especially interesting one is the ferroelectric smectic A phase (SmA$_{\text{F}}$) with spontaneous polarization along the layer normal observed in a few materials of the N$_{\text{F}}$ realm. Here, we present a mean-field molecular model that successfully captures the rich phase diagrams experimentally observed in the literature in terms of two parameters. Additionally, we carry out second harmonic generation, X-ray diffraction and birefringence measurements in a compound exhibiting the SmA$_{\text{F}}$ phase to determine the appropriate order parameters and compare with the model predictions.
\end{abstract}

\maketitle

\newpage

Mean-field theories constitute an intermediate approach between phenomenological and fully microscopic molecular theories. Even though they suffer from various limitations, mostly due to neglecting fluctuations, they sometimes provide valuable physical insights and have been applied to a wide range of systems \cite{book_phase_transitions, book_yeomans}. In the case of liquid crystals, Maier and Saupe proposed one such theory to describe the isotropic (Iso) to nematic (N) phase transition in 1960 \cite{maier_saupe_1, maier_saupe_2}. They assumed an anisotropic potential comprising steric and induced-dipole interactions in the hope that uniaxial nematic order would arise below a given temperature. This was indeed the case and, although the predicted behavior deviates somewhat from the experimental data, it provides an intuitive understanding of the Iso--N transition. Kobayashi later extended the Maier--Saupe theory to the smectic A (SmA) phase by introducing an additional order parameter related to the one-dimensional positional order along the nematic axis \cite{kobayashi_1, kobayashi_2}. It was McMillan, however, who studied the full consequences of introducing the possibility for such type of order, with its predictions again in qualitative agreement with experiment \cite{mcmillan}.

Currently, the liquid crystal community is witnessing a \textit{polar revolution} \cite{nf_review}. Aside from the fact that the long sought-after ferroelectric nematic (N$_{\text{F}}$) phase was experimentally realized in 2017 \cite{mandle_nematic_2017, nishikawa_fluid_2017, chen_first-principles_2020, Nishikawa_giant}, a great variety of related new polar phases have been found since. For instance, a splay-modulated antiferroelectric nematic phase, often called N$_{\text{S}}$ or SmZ$_{\text{A}}$, has been observed in some compounds directly preceding the N$_{\text{F}}$ phase on cooling \cite{ferroelastic, chen_smectic_za}. An especially interesting system is the ferroelectric smectic A (SmA$_{\text{F}}$) phase, for which the spontaneous polarization lies along the layer normal \cite{original_smaf, chen_smectic_a}. This phase is more viscous than the N$_{\text{F}}$ phase, and exhibits larger spontaneous polarization values. A visual sketch of the molecular arrangement in these phases is shown in Fig. S1. These recent findings have opened new avenues of exciting scientific research and promising technical applications. One of the most intriguing questions is how long-range polar order develops in these systems. Recently, an extension of the Maier-Saupe theory has been proposed by introducing a polar order parameter \cite{etxebarria_model}. The model successfully predicts the appearance of the N$_{\text{F}}$ phase, which can be reached from both the apolar Iso and N phases. In this Letter, we follow this procedure and generalize the model to include  the SmA$_{\text{F}}$ phase.

In the SmA phase, the molecules are preferentially oriented along the director $\mathbf{n}$ as in the N phase, but they also exhibit quasilong-range positional order in this direction forming a layered structure. Therefore, apart from the nematic order parameter $\eta_2 = \langle P_2(\cos{\theta}) \rangle$ ($\theta$ is the angle between the molecular long axis and $\mathbf{n}$, being $P_2(\cos{\theta})= (3\cos^2{\theta}-1)/2$ the second Legendre polynomial with the brackets denoting thermal averaging), one needs to introduce an additional order parameter $\sigma=\langle \cos{(qz)} P_2(\cos{\theta}) \rangle$, where $q=2\pi/d$, $d$ being the interplanar distance. Note that $\mathbf{n}$ and the layer normal have been taken along the $z$ direction. $\sigma$ thus describes the amplitude of the density wave along the director. No polar order parameter can be introduced since the SmA phase is apolar, i.e. the $\mathbf{n}$ and $-\mathbf{n}$ directions are equivalent. The anisotropic interaction between molecules is assumed to be short-ranged and proportional to $-\exp{\left[-(r_{12}/r_0)^2 \right]} P_2(\cos{\theta})$, where $r_{12}$ is the distance between centers of mass and $r_0$ is a measure of the interaction range. Within the mean-field approximation, it was shown that the one-particle potential can be written as

\begin{equation}
    V(z, \cos{\theta})=-V_2 \eta_2 P_2(\cos{\theta})-V_2 \alpha \sigma \cos{(qz)} P_2(\cos{\theta}),\label{eq:mcmillan}
\end{equation}
\noindent
where $V_2$ and $\alpha$ are constants with $V_2>0$ and $\alpha=2\exp{\left[-(\pi r_0/d)^2 \right]}$ \cite{mcmillan}. It is evident that $0<\alpha<2$, and $\alpha$ is larger for larger values of $d$ relative to $r_0$. Since $d$ is typically approximately a molecular length $l$, $\alpha$ increases with $l$.

As in the N$_{\text{F}}$ phase, in the SmA$_{\text{F}}$ phase the head-to-tail symmetry is broken, i.e. the $\mathbf{n}$ and $-\mathbf{n}$ directions are not equivalent. The simplest order parameter that can account for spontaneous polar order is $\eta_1=\langle P_1(\cos{\theta}) \rangle = \langle \cos{\theta} \rangle$. Therefore, as it was done in Ref. \onlinecite{etxebarria_model}, a natural extension of expression (\ref{eq:mcmillan}) to look for the SmA$_{\text{F}}$ phase is

\begin{equation}
    V(z, \cos{\theta})=-V_2 \eta_2 P_2(\cos{\theta})-V_2 \alpha \sigma \cos{(qz)}P_2(\cos{\theta}) -V_1 \eta_1 P_1(\cos{\theta}),
\end{equation}
\noindent
where $V_1>0$ is a new constant measuring the strength of the polar interaction. Note that the polar and mesogenic interactions are decoupled. With this potential, the probability of a molecule being located at a distance $z$ from a smectic plane at the origin and oriented at an angle $\theta$ from the director is given by

\begin{equation}
    \mathcal{P}(z, \theta)=\frac{1}{Z}\exp{\left[\frac{V_2 \eta_2 P_2(\cos{\theta})+V_2 \alpha \sigma \cos{(qz)}P_2(\cos{\theta}) +V_1 \eta_1 P_1(\cos{\theta})}{k_{\text{B}} T} \right]},
\end{equation}
\noindent
where $k_{\text{B}}$ is the Boltzmann constant, $T$ the temperature, and $Z$ the partition function:

\begin{equation}
    Z=\int_{0}^{d} \int_{0}^{\pi}\exp{\left[\frac{V_2 \eta_2 P_2(\cos{\theta})+V_2 \alpha \sigma \cos{(qz)}P_2(\cos{\theta}) +V_1 \eta_1 P_1(\cos{\theta})}{k_{\text{B}} T} \right]} \sin{\theta} \text{d}\theta \text{d}z.
\end{equation}

Accordingly, the following self-consistent system of equations can be solved in order to obtain the order parameters at each temperature:

\begin{eqnarray}
    \eta_1=\frac{1}{Z}\int_{0}^{d} \int_{0}^{\pi}P_1(\cos{\theta}) \mathcal{P}(z, \theta) \sin{\theta} \text{d}\theta \text{d}z; \label{eta1} \\
    \eta_2=\frac{1}{Z}\int_{0}^{d} \int_{0}^{\pi}P_2(\cos{\theta}) \mathcal{P}(z, \theta) \sin{\theta} \text{d}\theta \text{d}z; \label{eta2} \\
    \sigma=\frac{1}{Z}\int_{0}^{d} \int_{0}^{\pi}\cos{(qz)}P_2(\cos{\theta})\mathcal{P}(z, \theta) \sin{\theta} \text{d}\theta \text{d}z.\label{sigma}
\end{eqnarray}

The order parameters can also be calculated by minimizing the free energy at a given temperature. Considering that the internal energy of a system of $N$ molecules is $U=N \langle V \rangle /2$ and the entropy is $S=-N k_{\text{B}}\langle \ln{\mathcal{P}} \rangle$, the Helmholtz free energy can be easily calculated as $F=U-TS$. The end result is:

\begin{equation}
    F/N=\frac{1}{2}V_2 \eta_2^2+\frac{1}{2}V_2 \alpha \sigma^2 +\frac{1}{2}V_1 \eta_1^2-k_{\text{B}}T\ln{Z}.\label{free_energy}
\end{equation}

In our case, we have numerically minimized Eq. (\ref{free_energy}) to obtain the order parameters at each temperature, later checking that Eqs. (\ref{eta1}), (\ref{eta2}) and (\ref{sigma}) are satisfied.

Aside from the order parameters, of course, the thermal average of any other physical quantity of interest can be computed in an analogous manner. For instance, the SmA$_{\text{F}}$ phase shows nonlinear optical activity. Since it has the same point group symmetry $C_{\infty \text{v}}$ as the N$_{\text{F}}$ phase, its second order dielectric susceptibility tensor has the same form as in Ref. \onlinecite{folcia_ferroelectric_2022}, where it was shown that the optical second-harmonic generation intensity must be proportional to $\langle \cos^3{\theta} \rangle^2$ in a first approximation. Another physical quantity of great interest when it comes to phase transitions is the heat capacity, which at constant volume can be obtained as $C_V=T \left(\partial S/\partial T \right)$. Lastly, the entropy change at a transition can be written as

\begin{equation}
    \Delta S/N k_{\text{B}} = -\frac{V_2}{k_{\text{B}}T}\Delta \eta_2^2 - \frac{V_2 \alpha}{k_{\text{B}}T}\Delta \sigma^2 - \frac{V_1}{k_{\text{B}}T}\Delta \eta_1^2 + \Delta \ln{Z}.
\end{equation}

\begin{figure}
\includegraphics[width=0.5\textwidth]{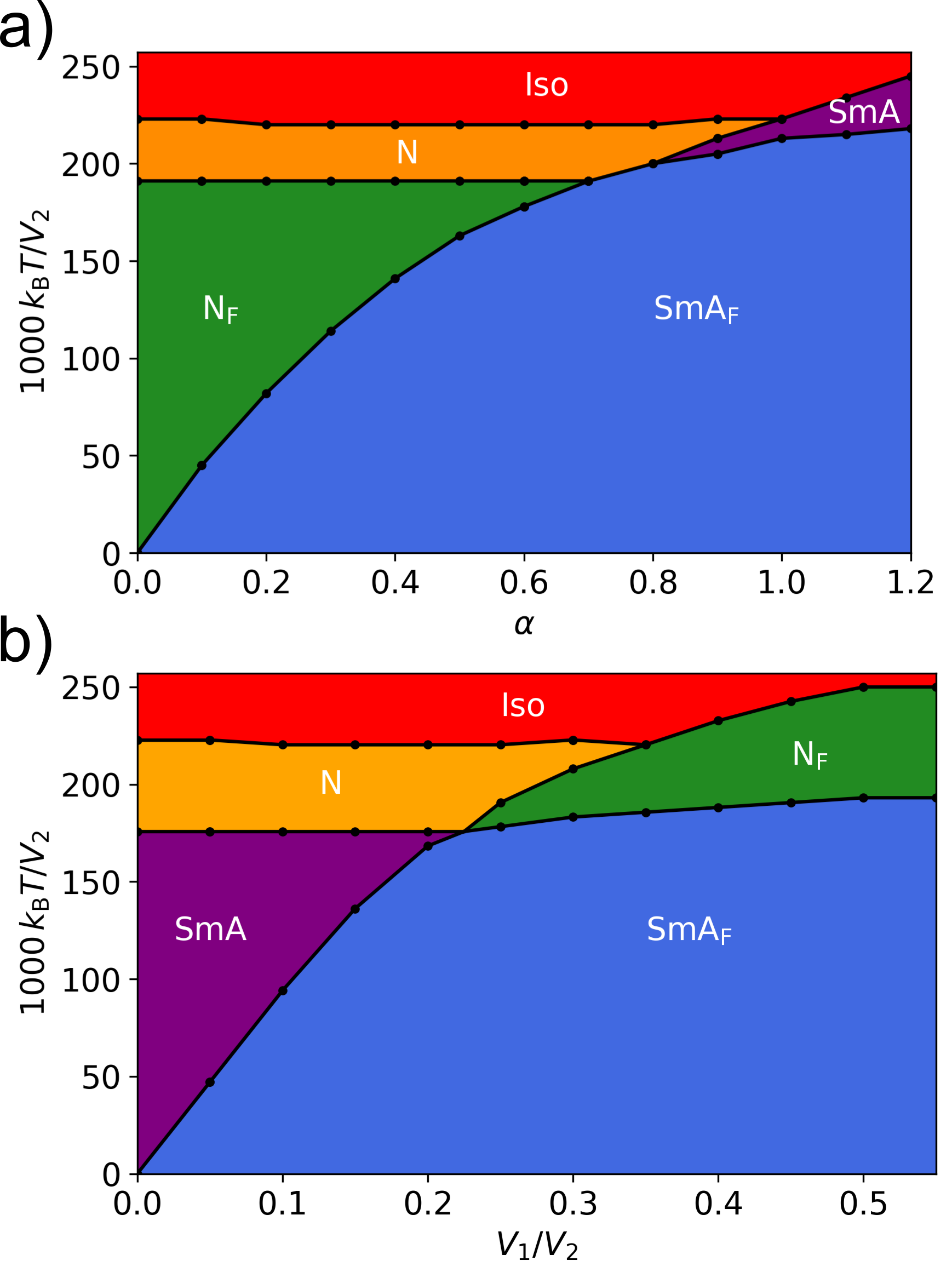}
\caption{\label{fig:main_phase_diagrams} (a) Predicted phase diagram for $V_1/V_2=0.25$ as a function of the parameter $\alpha$. (b) Predicted phase diagram for $\alpha=0.6$ as a function of the parameter $V_1/V_2$.}
\end{figure}

Firstly, we present in Fig. \ref{fig:main_phase_diagrams} a couple of phase diagrams that we consider to be representative of the model's predictions. Fig. \ref{fig:main_phase_diagrams}a shows the different phase sequences that are possible for a given value (0.25) of the ratio $V_1/V_2$, which tunes the strength of the polar interactions, as a function of the parameter $\alpha$. For small values of $\alpha$ we observe the Iso--N--N$_{\text{F}}$--SmA$_{\text{F}}$ phase sequence. As expected, the transition temperature to the SmA$_{\text{F}}$ phase increases with $\alpha$. At larger values of $\alpha$ there is a narrow interval where Iso--N--SmA$_{\text{F}}$ and Iso--N--SmA--SmA$_{\text{F}}$ can occur. For $\alpha>1$, purely smectic order is favored, allowing the Iso--SmA--SmA$_{\text{F}}$ phase sequence. It is also instructive to see how the phase diagram changes as a function of $V_1/V_2$ for constant $\alpha$, as shown in Fig. \ref{fig:main_phase_diagrams}b for $\alpha=0.6$. If the strength of the polar interaction is small, i.e. for small $V_1/V_2$, the Iso--N--SmA--SmA$_{\text{F}}$ phase sequence is observed. Again, as expected, the larger the ratio $V_1/V_2$, the higher the transition temperature to the SmA$_{\text{F}}$ phase. In fact, it increases quite rapidly at first and then saturates. As the polar interaction becomes stronger, the N$_{\text{F}}$ phase enters the picture, thus obtaining the Iso--N--N$_{\text{F}}$--SmA$_{\text{F}}$ and even the Iso--N$_{\text{F}}$--SmA$_{\text{F}}$ phase sequences. It should be mentioned that the apparent quadruple point in this phase diagram is not such, but rather the result of the limited resolution of the calculations. Lastly, it is important to point out that the order of the transitions changes with the model parameters.

\begin{figure}
\includegraphics[width=0.5\textwidth]{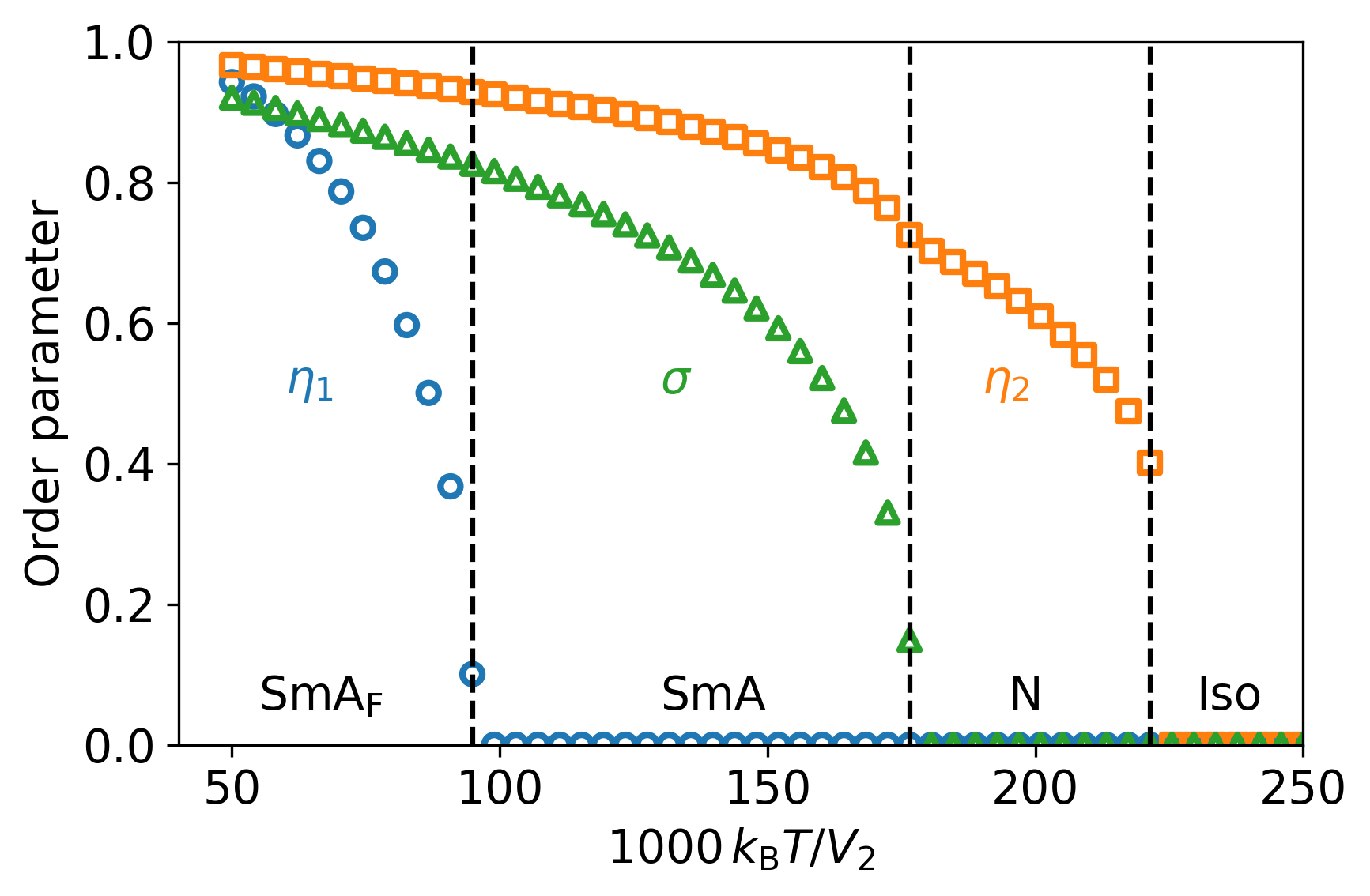}
\caption{\label{fig:mandle_compound_1} Temperature dependence of the order parameters $\eta_1$ (blue open circles), $\eta_2$ (orange open squares) and $\sigma$ (green open triangles) for an Iso--N--SmA--SmA$_{\text{F}}$ phase sequence with $\alpha=0.6$ and $V_1/V_2=0.1$.}
\end{figure}

In order to check the validity of the model to describe the aforementioned phase transitions, we will analyze the temperature behavior of the order parameters and compare it to available experimental measurements. For instance, it is well-known that the spontaneous polarization is proportional to $\eta_1$, and that the birefringence is proportional to $\eta_2$. In Ref. \onlinecite{mandle_compound_1}, Gibb et al. report a material (compound $\mathbf{1}$) that exhibits an Iso--N--SmA--SmA$_{\text{F}}$ phase sequence (plus a low-temperature SmC$^{\text{P}}_{\text{H}}$ which is out of the scope of this paper), for which good-quality measurements were obtained. Choosing $\alpha=0.6$ and $V_1/V_2=0.1$ we can reproduce this phase sequence. The predicted evolution of the order parameters is shown in Fig. \ref{fig:mandle_compound_1}. At the Iso--N phase transition, $\eta_2$ exhibits a jump due to the sudden appearance of orientational order, which is always first-order. The N--SmA phase transition is also first-order and a jump in $\sigma$ is observed, as expected. It is also manifested as a small jump in $\eta_2$, which bears a strong resemblance to the measured birefringence curve (see Fig. 2d in Ref. \onlinecite{mandle_compound_1}). When passing to the SmA$_{\text{F}}$ phase, $\eta_1$ becomes-non zero, and the order parameters (and entropy) exhibit no discontinuity due to it being a second-order phase transition. This is also close to reality, since the transition is weakly first-order. As the temperature is lowered in the SmA$_{\text{F}}$ phase, measurements show a steeper increase in birefringence, but it could also be related to pretransitional effects close to a different lower-temperature phase. In any case, the authors directly measured $\eta_2$ in this temperature interval via polarized Raman spectroscopy, staying practically constant in the polar phase (Fig. 2e in their paper), as predicted by our model.

In a subsequent paper, having studied multiple pure compounds and mixtures, the same group of authors arrives at the conclusion that the mechanisms driving polar order are more or less independent of the mechanisms that drive mesogenic order \cite{mandle_hobbs}. Therefore, they argue that transitions that involve a change in both mesogenic and polar order are improbable, e.g. a N--SmA$_{\text{F}}$ phase transition. This transition is, however, possible, as could already be observed in Fig. \ref{fig:main_phase_diagrams}, and it has been recently realized in compound EST-4 in Ref. \onlinecite{kikuchi_n_smaf}. As measurements, the authors only present a DSC curve which qualitatively agrees with our calculations (see Fig. S2).

\begin{figure}
\includegraphics[width=0.5\textwidth]{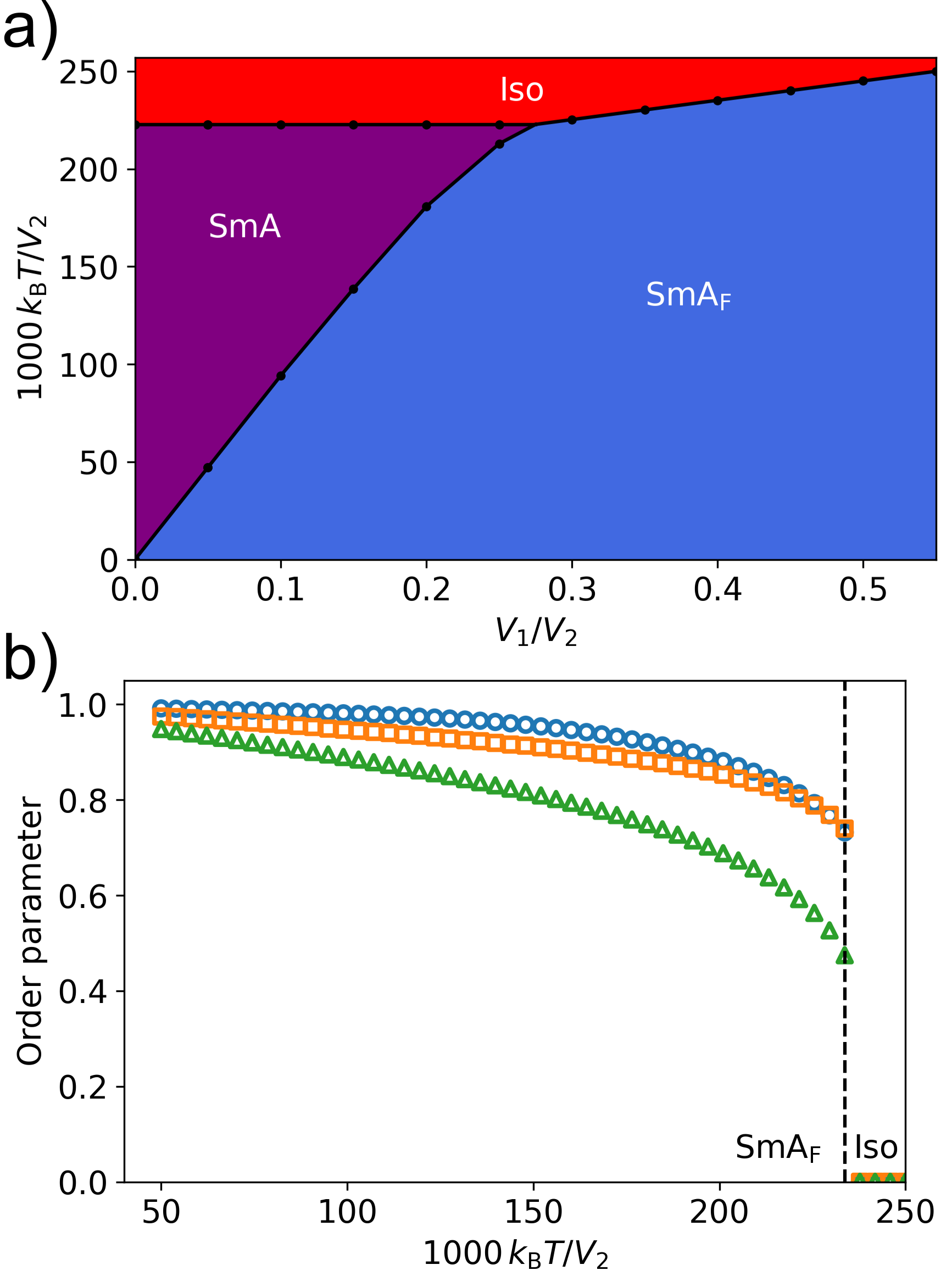}
\caption{\label{fig:iso_smaf_final}(a) Predicted phase diagram for $\alpha=1$ as a function of the parameter $V_1/V_2$. (b) Temperature dependence of the order parameters $\eta_1$ (blue open circles), $\eta_2$ (orange open squares) and $\sigma$ (green open triangles) for an Iso--SmA$_{\text{F}}$ phase sequence with $\alpha=1$ and $V_1/V_2=0.4$.}
\end{figure}

Another improbable transition would be a direct Iso--SmA$_{\text{F}}$ transition. This implies that, at a certain temperature, orientational, positional and polar order must be favored at the same time, which seems unfeasible. In the case of polar nematics, however, a direct Iso--N$_{\text{F}}$ transition was found a few years ago \cite{manabe_original}. In terms of our model, we can expect an Iso--SmA$_{\text{F}}$ transition at large values of both $\alpha$ and $V_1/V_2$. Consequently, we have plotted in Fig. \ref{fig:iso_smaf_final}a a theoretical phase diagram for $\alpha=1$ as a function of $V_1/V_2$. As expected, for small values of $V_1/V_2$ we get an Iso--SmA--SmA$_{\text{F}}$ phase sequence. As the strength of the polar interaction increases, the temperature interval of the SmA phase becomes narrower until it disappears. For $V_1/V_2 \geq 0.28$ a direct Iso--SmA$_{\text{F}}$ phase transition is possible. These results imply that the molecule needs to be long and, at the same time, the interaction that gives rise to polar order needs to be strong. In Fig. \ref{fig:iso_smaf_final}b we can see what the evolution of the order parameters might look like for a particular case with $\alpha=1$ and $V_1/V_2=0.4$. Since this is a strongly first-order transition, a larger $|\Delta S|$ is to be expected in comparison with the previous transitions. Indeed, with the mentioned parameters $|\Delta S|=16$ J/mol K. Interestingly, a compound exhibiting this kind of phase transition has recently been synthesized (compound \textbf{10f} in Ref. \onlinecite{iso_smaf}). The transition occurs at $148^{\circ}$C, and the measured latent heat is $8.5$ kJ/mol, so that $|\Delta S|=20.2$ J/mol K, again in qualitative agreement with the theoretical value. The authors also provide spontaneous polarization measurements as a function of temperature in Fig. S25. Even though there are only a few points near the phase transition temperature, the curve points to a sudden increase of polar order upon cooling from the Iso phase, which rapidly saturates to its maximum value, as predicted in Fig. \ref{fig:iso_smaf_final}.

Lastly, we will present the results of our own experiments to make a more direct comparison with the model's predictions. The studied compound is DIO-CN, an analog of the prototypical ferroelectric nematogen DIO. Details about the synthesis of this material can be found in Ref. \onlinecite{diocn}. It is a highly polar molecule, with a dipole moment of $14.6$ D \cite{diocn}, and exhibits an Iso--$233^{\circ}$C--N--$190^{\circ}$C--N$_{\text{S}}$--$175^{\circ}$C--N$_{\text{F}}$--$115^{\circ}$C--SmA$_{\text{F}}$--$65^{\circ}$C--Cry phase sequence. The appearence of the N$_{\text{S}}$ phase is out of the scope of this paper but, since its temperature interval is narrow compared to the other phases, it will not hamper our analysis. In particular, we performed second-harmonic generation (SHG), birefringence and X-ray diffraction (XRD) measurements as a function of temperature (experimental details in the Supplemental Material). In Fig. \ref{fig:diocn_results} we have plotted the birefringence and SHG experimental results alongside with the theoretically predicted curves for model parameters $\alpha=0.55$ and $V_1/V_2=0.3$. As can be seen in Fig. \ref{fig:diocn_results}a, the material shows no nonlinear optical activity in the N and N$_{\text{S}}$ phases, as expected. At the transition to the N$_{\text{F}}$ phase, however, the SHG intensity suddenly increases due to the emergence of long-range ferroelectric order. The reason why the experimental behavior is steeper than the theoretical one in Fig. \ref{fig:diocn_results}b is because the SHG intensity, aside from depending on the second order dielectric susceptibility, carries a $\sinc^2{(\Delta k \,l/2)}$ term, which depends on the dispersion of the extraordinary index ($l$ being the sample thickness). This quantity can exhibit a jump at the transition that explains a larger jump in the SHG signal. The SHG intensity continues increasing down to the N$_{\text{F}}$--SmA$_{\text{F}}$ transition, where it exhibits a small bump. In this region, the $\sinc^2{(\Delta k \,l/2)}$ term is expected to be approximately constant, and the temperature dependence of the SHG signal is essentially due to $\langle \cos^3{\theta} \rangle ^2$. With regards to the birefringence $\Delta n$, as shown in Fig. \ref{fig:diocn_results}a, it exhibits an anomaly which is correctly captured by the model (Fig. \ref{fig:diocn_results}b). Furthermore, the model also predicts a slightly larger jump at the N--N$_{\text{F}}$ transition, which was already observed in Ref. \onlinecite{diocn} (although the N$_{\text{S}}$ phase appears in the middle). We also performed XRD measurements and integrated the peak intensity of the (001) smectic plane to verify its proportionality to $\sigma$ \cite{mcmillan} (see Figs. S3-S4). 


\begin{figure}
\includegraphics[width=0.5\textwidth]{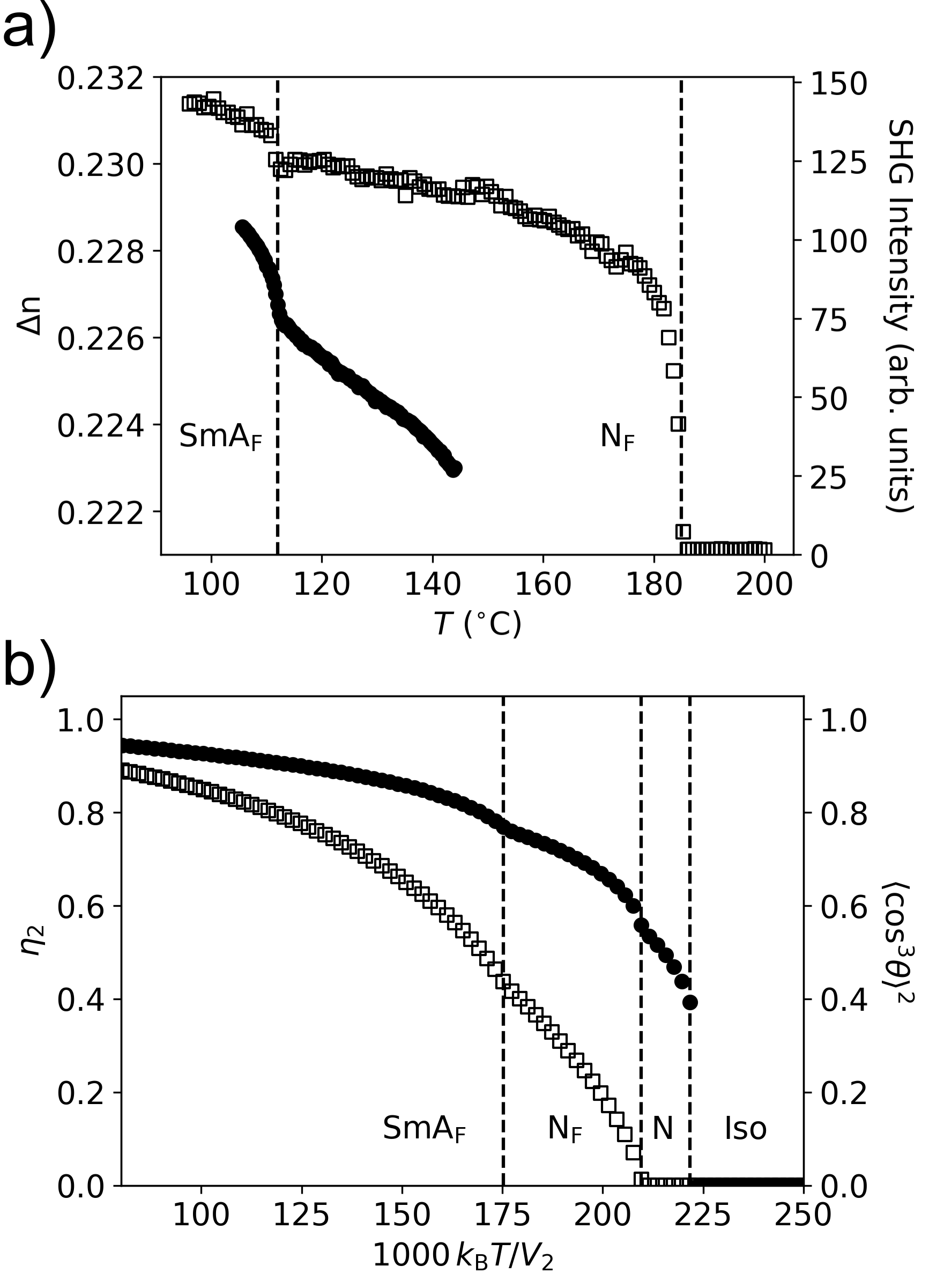}
\caption{\label{fig:diocn_results} (a) Experimental birefringence (solid circles) and SHG intensity (empty squares) as a function of temperature for DIO-CN. (b) Corresponding theoretical predictions with parameters $\alpha=0.55$ and $V_1/V_2=0.3$.}
\end{figure}

In summary, the model successfully predicts a wide range of phase sequences and provides a simple intuitive picture of why certain phases are favored over others as a result of the interplay between two material parameters. It is important to emphasize that we are still in the process of understanding how polar order emerges in these systems. For this purpose, a fully microscopic theory that focuses on the specific mechanism that makes the parallel alignment of dipoles favorable should be sought after. One example in this category is the mechanism proposed by Madhusudana for the N$_{\text{F}}$ phase \cite{madhusudana} who studied the electrostatic interaction between molecules modeled as charge density waves and found that the parallel alignment of dipoles is favored. Although this is definitely a promising result, the model has some shortcomings. For example, it has been shown that if the number of charge density waves is slightly reduced, to better represent some ferroelectric nematogens, the antiparallel alignment is preferred \cite{erkoreka_uuu}. Moreover, as recently noted by Osipov, Madhusudana's results are insufficient, because it is also important to analyze if such an interaction contributes negatively to the free energy so as to overcome the orientational entropy \cite{osipov, osipov_review}. Osipov has also proposed a model combining dipole-dipole interactions and short-range orientational-translational correlations. Finally, a promising density functional theory approach focusing on dipole-dipole interactions has been proposed \cite{dipoles_nf}. Further research in this and other directions should be undertaken to arrive at a consistent theory of these novel polar phases of the N$_{\text{F}}$ realm.

\begin{acknowledgments}
We thank J. Etxebarria, J. Ortega and C.L. Folcia for the careful reading of the manuscript. A.E. and J.M.-P. acknowledge funding from the Basque Government Project IT1458-22. A.E. thanks the Department of Education of the Basque Government for a predoctoral fellowship (grant no. PRE\_2023\_2\_0113). S.A. and M.H. acknowledge the funding supports from the National Key Research and Development Program of China (No. 2022YFA1405000), and Natural Science Foundation of Guangdong Province (2024B1515040023).
\end{acknowledgments}

\bibliography{REFERENCES}

\begin{thebibliography}{29}%
\makeatletter
\providecommand \@ifxundefined [1]{%
 \@ifx{#1\undefined}
}%
\providecommand \@ifnum [1]{%
 \ifnum #1\expandafter \@firstoftwo
 \else \expandafter \@secondoftwo
 \fi
}%
\providecommand \@ifx [1]{%
 \ifx #1\expandafter \@firstoftwo
 \else \expandafter \@secondoftwo
 \fi
}%
\providecommand \natexlab [1]{#1}%
\providecommand \enquote  [1]{``#1''}%
\providecommand \bibnamefont  [1]{#1}%
\providecommand \bibfnamefont [1]{#1}%
\providecommand \citenamefont [1]{#1}%
\providecommand \href@noop [0]{\@secondoftwo}%
\providecommand \href [0]{\begingroup \@sanitize@url \@href}%
\providecommand \@href[1]{\@@startlink{#1}\@@href}%
\providecommand \@@href[1]{\endgroup#1\@@endlink}%
\providecommand \@sanitize@url [0]{\catcode `\\12\catcode `\$12\catcode `\&12\catcode `\#12\catcode `\^12\catcode `\_12\catcode `\%12\relax}%
\providecommand \@@startlink[1]{}%
\providecommand \@@endlink[0]{}%
\providecommand \url  [0]{\begingroup\@sanitize@url \@url }%
\providecommand \@url [1]{\endgroup\@href {#1}{\urlprefix }}%
\providecommand \urlprefix  [0]{URL }%
\providecommand \Eprint [0]{\href }%
\providecommand \doibase [0]{http://dx.doi.org/}%
\providecommand \selectlanguage [0]{\@gobble}%
\providecommand \bibinfo  [0]{\@secondoftwo}%
\providecommand \bibfield  [0]{\@secondoftwo}%
\providecommand \translation [1]{[#1]}%
\providecommand \BibitemOpen [0]{}%
\providecommand \bibitemStop [0]{}%
\providecommand \bibitemNoStop [0]{.\EOS\space}%
\providecommand \EOS [0]{\spacefactor3000\relax}%
\providecommand \BibitemShut  [1]{\csname bibitem#1\endcsname}%
\let\auto@bib@innerbib\@empty
\bibitem [{\citenamefont {Papon}\ \emph {et~al.}(2002)\citenamefont {Papon}, \citenamefont {Leblond},\ and\ \citenamefont {Meijer}}]{book_phase_transitions}%
  \BibitemOpen
  \bibfield  {author} {\bibinfo {author} {\bibfnamefont {P.}~\bibnamefont {Papon}}, \bibinfo {author} {\bibfnamefont {J.}~\bibnamefont {Leblond}}, \ and\ \bibinfo {author} {\bibfnamefont {P.~H.~E.}\ \bibnamefont {Meijer}},\ }\href {\doibase 10.1007/978-3-662-04989-1} {\emph {\bibinfo {title} {The Physics of Phase Transitions}}}\ (\bibinfo  {publisher} {Springer Berlin Heidelberg},\ \bibinfo {year} {2002})\BibitemShut {NoStop}%
\bibitem [{\citenamefont {Yeomans}(1992)}]{book_yeomans}%
  \BibitemOpen
  \bibfield  {author} {\bibinfo {author} {\bibfnamefont {J.~M.}\ \bibnamefont {Yeomans}},\ }\href@noop {} {\emph {\bibinfo {title} {Statistical mechanics of phase transitions}}}\ (\bibinfo  {publisher} {Clarendon Press},\ \bibinfo {address} {Oxford, England},\ \bibinfo {year} {1992})\BibitemShut {NoStop}%
\bibitem [{\citenamefont {Maier}\ and\ \citenamefont {Saupe}(1958)}]{maier_saupe_1}%
  \BibitemOpen
  \bibfield  {author} {\bibinfo {author} {\bibfnamefont {W.}~\bibnamefont {Maier}}\ and\ \bibinfo {author} {\bibfnamefont {A.}~\bibnamefont {Saupe}},\ }\href {\doibase 10.1515/zna-1958-0716} {\bibfield  {journal} {\bibinfo  {journal} {Zeitschrift f\"{u}r Naturforschung A}\ }\textbf {\bibinfo {volume} {13}},\ \bibinfo {pages} {564–566} (\bibinfo {year} {1958})}\BibitemShut {NoStop}%
\bibitem [{\citenamefont {Maier}\ and\ \citenamefont {Saupe}(1960)}]{maier_saupe_2}%
  \BibitemOpen
  \bibfield  {author} {\bibinfo {author} {\bibfnamefont {W.}~\bibnamefont {Maier}}\ and\ \bibinfo {author} {\bibfnamefont {A.}~\bibnamefont {Saupe}},\ }\href {\doibase 10.1515/zna-1960-0401} {\bibfield  {journal} {\bibinfo  {journal} {Zeitschrift f\"{u}r Naturforschung A}\ }\textbf {\bibinfo {volume} {15}},\ \bibinfo {pages} {287–292} (\bibinfo {year} {1960})}\BibitemShut {NoStop}%
\bibitem [{\citenamefont {Kobayashi}(1970)}]{kobayashi_1}%
  \BibitemOpen
  \bibfield  {author} {\bibinfo {author} {\bibfnamefont {K.}~\bibnamefont {Kobayashi}},\ }\href {\doibase 10.1016/0375-9601(70)90186-6} {\bibfield  {journal} {\bibinfo  {journal} {Physics Letters A}\ }\textbf {\bibinfo {volume} {31}},\ \bibinfo {pages} {125–126} (\bibinfo {year} {1970})}\BibitemShut {NoStop}%
\bibitem [{\citenamefont {K.~Kobayashi}(1970)}]{kobayashi_2}%
  \BibitemOpen
  \bibfield  {author} {\bibinfo {author} {\bibfnamefont {K.}~\bibnamefont {K.~Kobayashi}},\ }\href {\doibase 10.1143/jpsj.29.101} {\bibfield  {journal} {\bibinfo  {journal} {Journal of the Physical Society of Japan}\ }\textbf {\bibinfo {volume} {29}},\ \bibinfo {pages} {101–105} (\bibinfo {year} {1970})}\BibitemShut {NoStop}%
\bibitem [{\citenamefont {McMillan}(1971)}]{mcmillan}%
  \BibitemOpen
  \bibfield  {author} {\bibinfo {author} {\bibfnamefont {W.~L.}\ \bibnamefont {McMillan}},\ }\href {\doibase 10.1103/physreva.4.1238} {\bibfield  {journal} {\bibinfo  {journal} {Physical Review A}\ }\textbf {\bibinfo {volume} {4}},\ \bibinfo {pages} {1238–1246} (\bibinfo {year} {1971})}\BibitemShut {NoStop}%
\bibitem [{\citenamefont {Sebasti\'an}\ \emph {et~al.}(2022)\citenamefont {Sebasti\'an}, \citenamefont {\ifmmode \check{C}\else \v{C}\fi{}opi\ifmmode~\check{c}\else \v{c}\fi{}},\ and\ \citenamefont {Mertelj}}]{nf_review}%
  \BibitemOpen
  \bibfield  {author} {\bibinfo {author} {\bibfnamefont {N.}~\bibnamefont {Sebasti\'an}}, \bibinfo {author} {\bibfnamefont {M.}~\bibnamefont {\ifmmode \check{C}\else \v{C}\fi{}opi\ifmmode~\check{c}\else \v{c}\fi{}}}, \ and\ \bibinfo {author} {\bibfnamefont {A.}~\bibnamefont {Mertelj}},\ }\href {\doibase 10.1103/PhysRevE.106.021001} {\bibfield  {journal} {\bibinfo  {journal} {Physical Review E}\ }\textbf {\bibinfo {volume} {106}},\ \bibinfo {pages} {021001} (\bibinfo {year} {2022})}\BibitemShut {NoStop}%
\bibitem [{\citenamefont {Mandle}\ \emph {et~al.}(2017)\citenamefont {Mandle}, \citenamefont {Cowling},\ and\ \citenamefont {Goodby}}]{mandle_nematic_2017}%
  \BibitemOpen
  \bibfield  {author} {\bibinfo {author} {\bibfnamefont {R.~J.}\ \bibnamefont {Mandle}}, \bibinfo {author} {\bibfnamefont {S.~J.}\ \bibnamefont {Cowling}}, \ and\ \bibinfo {author} {\bibfnamefont {J.~W.}\ \bibnamefont {Goodby}},\ }\href {\doibase 10.1039/C7CP00456G} {\bibfield  {journal} {\bibinfo  {journal} {Physical Chemistry Chemical Physics}\ }\textbf {\bibinfo {volume} {19}},\ \bibinfo {pages} {11429} (\bibinfo {year} {2017})}\BibitemShut {NoStop}%
\bibitem [{\citenamefont {Nishikawa}\ \emph {et~al.}(2017)\citenamefont {Nishikawa}, \citenamefont {Shiroshita}, \citenamefont {Higuchi}, \citenamefont {Okumura}, \citenamefont {Haseba}, \citenamefont {Yamamoto}, \citenamefont {Sago},\ and\ \citenamefont {Kikuchi}}]{nishikawa_fluid_2017}%
  \BibitemOpen
  \bibfield  {author} {\bibinfo {author} {\bibfnamefont {H.}~\bibnamefont {Nishikawa}}, \bibinfo {author} {\bibfnamefont {K.}~\bibnamefont {Shiroshita}}, \bibinfo {author} {\bibfnamefont {H.}~\bibnamefont {Higuchi}}, \bibinfo {author} {\bibfnamefont {Y.}~\bibnamefont {Okumura}}, \bibinfo {author} {\bibfnamefont {Y.}~\bibnamefont {Haseba}}, \bibinfo {author} {\bibfnamefont {S.-i.}\ \bibnamefont {Yamamoto}}, \bibinfo {author} {\bibfnamefont {K.}~\bibnamefont {Sago}}, \ and\ \bibinfo {author} {\bibfnamefont {H.}~\bibnamefont {Kikuchi}},\ }\href {\doibase 10.1002/adma.201702354} {\bibfield  {journal} {\bibinfo  {journal} {Advanced Materials}\ }\textbf {\bibinfo {volume} {29}},\ \bibinfo {pages} {1702354} (\bibinfo {year} {2017})}\BibitemShut {NoStop}%
\bibitem [{\citenamefont {Chen}\ \emph {et~al.}(2020)\citenamefont {Chen}, \citenamefont {Korblova}, \citenamefont {Dong}, \citenamefont {Wei}, \citenamefont {Shao}, \citenamefont {Radzihovsky}, \citenamefont {Glaser}, \citenamefont {Maclennan}, \citenamefont {Bedrov}, \citenamefont {Walba},\ and\ \citenamefont {Clark}}]{chen_first-principles_2020}%
  \BibitemOpen
  \bibfield  {author} {\bibinfo {author} {\bibfnamefont {X.}~\bibnamefont {Chen}}, \bibinfo {author} {\bibfnamefont {E.}~\bibnamefont {Korblova}}, \bibinfo {author} {\bibfnamefont {D.}~\bibnamefont {Dong}}, \bibinfo {author} {\bibfnamefont {X.}~\bibnamefont {Wei}}, \bibinfo {author} {\bibfnamefont {R.}~\bibnamefont {Shao}}, \bibinfo {author} {\bibfnamefont {L.}~\bibnamefont {Radzihovsky}}, \bibinfo {author} {\bibfnamefont {M.~A.}\ \bibnamefont {Glaser}}, \bibinfo {author} {\bibfnamefont {J.~E.}\ \bibnamefont {Maclennan}}, \bibinfo {author} {\bibfnamefont {D.}~\bibnamefont {Bedrov}}, \bibinfo {author} {\bibfnamefont {D.~M.}\ \bibnamefont {Walba}}, \ and\ \bibinfo {author} {\bibfnamefont {N.~A.}\ \bibnamefont {Clark}},\ }\href {\doibase 10.1073/pnas.2002290117} {\bibfield  {journal} {\bibinfo  {journal} {Proceedings of the National Academy of Sciences}\ }\textbf {\bibinfo {volume} {117}},\ \bibinfo {pages} {14021} (\bibinfo {year} {2020})}\BibitemShut {NoStop}%
\bibitem [{\citenamefont {Li}\ \emph {et~al.}(2021)\citenamefont {Li}, \citenamefont {Nishikawa}, \citenamefont {Kougo}, \citenamefont {Zhou}, \citenamefont {Dai}, \citenamefont {Tang}, \citenamefont {Zhao}, \citenamefont {Hisai}, \citenamefont {Huang},\ and\ \citenamefont {Aya}}]{Nishikawa_giant}%
  \BibitemOpen
  \bibfield  {author} {\bibinfo {author} {\bibfnamefont {J.}~\bibnamefont {Li}}, \bibinfo {author} {\bibfnamefont {H.}~\bibnamefont {Nishikawa}}, \bibinfo {author} {\bibfnamefont {J.}~\bibnamefont {Kougo}}, \bibinfo {author} {\bibfnamefont {J.}~\bibnamefont {Zhou}}, \bibinfo {author} {\bibfnamefont {S.}~\bibnamefont {Dai}}, \bibinfo {author} {\bibfnamefont {W.}~\bibnamefont {Tang}}, \bibinfo {author} {\bibfnamefont {X.}~\bibnamefont {Zhao}}, \bibinfo {author} {\bibfnamefont {Y.}~\bibnamefont {Hisai}}, \bibinfo {author} {\bibfnamefont {M.}~\bibnamefont {Huang}}, \ and\ \bibinfo {author} {\bibfnamefont {S.}~\bibnamefont {Aya}},\ }\href {\doibase 10.1126/sciadv.abf5047} {\bibfield  {journal} {\bibinfo  {journal} {Science Advances}\ }\textbf {\bibinfo {volume} {7}},\ \bibinfo {pages} {eabf5047} (\bibinfo {year} {2021})}\BibitemShut {NoStop}%
\bibitem [{\citenamefont {Sebasti\'an}\ \emph {et~al.}(2020)\citenamefont {Sebasti\'an}, \citenamefont {Cmok}, \citenamefont {Mandle}, \citenamefont {de~la Fuente}, \citenamefont {Dreven\ifmmode\check{s}\else\v{s}\fi{}ek~Olenik}, \citenamefont {\ifmmode \check{C}\else \v{C}\fi{}opi\ifmmode~\check{c}\else \v{c}\fi{}},\ and\ \citenamefont {Mertelj}}]{ferroelastic}%
  \BibitemOpen
  \bibfield  {author} {\bibinfo {author} {\bibfnamefont {N.}~\bibnamefont {Sebasti\'an}}, \bibinfo {author} {\bibfnamefont {L.}~\bibnamefont {Cmok}}, \bibinfo {author} {\bibfnamefont {R.~J.}\ \bibnamefont {Mandle}}, \bibinfo {author} {\bibfnamefont {M.~R.}\ \bibnamefont {de~la Fuente}}, \bibinfo {author} {\bibfnamefont {I.}~\bibnamefont {Dreven\ifmmode\check{s}\else\v{s}\fi{}ek~Olenik}}, \bibinfo {author} {\bibfnamefont {M.}~\bibnamefont {\ifmmode \check{C}\else \v{C}\fi{}opi\ifmmode~\check{c}\else \v{c}\fi{}}}, \ and\ \bibinfo {author} {\bibfnamefont {A.}~\bibnamefont {Mertelj}},\ }\href {\doibase 10.1103/PhysRevLett.124.037801} {\bibfield  {journal} {\bibinfo  {journal} {Physical Review Letters}\ }\textbf {\bibinfo {volume} {124}},\ \bibinfo {pages} {037801} (\bibinfo {year} {2020})}\BibitemShut {NoStop}%
\bibitem [{\citenamefont {Chen}\ \emph {et~al.}(2023)\citenamefont {Chen}, \citenamefont {Martinez}, \citenamefont {Korblova}, \citenamefont {Freychet}, \citenamefont {Zhernenkov}, \citenamefont {Glaser}, \citenamefont {Wang}, \citenamefont {Zhu}, \citenamefont {Radzihovsky}, \citenamefont {Maclennan}, \citenamefont {Walba},\ and\ \citenamefont {Clark}}]{chen_smectic_za}%
  \BibitemOpen
  \bibfield  {author} {\bibinfo {author} {\bibfnamefont {X.}~\bibnamefont {Chen}}, \bibinfo {author} {\bibfnamefont {V.}~\bibnamefont {Martinez}}, \bibinfo {author} {\bibfnamefont {E.}~\bibnamefont {Korblova}}, \bibinfo {author} {\bibfnamefont {G.}~\bibnamefont {Freychet}}, \bibinfo {author} {\bibfnamefont {M.}~\bibnamefont {Zhernenkov}}, \bibinfo {author} {\bibfnamefont {M.~A.}\ \bibnamefont {Glaser}}, \bibinfo {author} {\bibfnamefont {C.}~\bibnamefont {Wang}}, \bibinfo {author} {\bibfnamefont {C.}~\bibnamefont {Zhu}}, \bibinfo {author} {\bibfnamefont {L.}~\bibnamefont {Radzihovsky}}, \bibinfo {author} {\bibfnamefont {J.~E.}\ \bibnamefont {Maclennan}}, \bibinfo {author} {\bibfnamefont {D.~M.}\ \bibnamefont {Walba}}, \ and\ \bibinfo {author} {\bibfnamefont {N.~A.}\ \bibnamefont {Clark}},\ }\href {\doibase 10.1073/pnas.2217150120} {\bibfield  {journal} {\bibinfo  {journal} {Proceedings of the National Academy of Sciences}\ }\textbf {\bibinfo {volume} {120}},\ \bibinfo {pages} {e2217150120} (\bibinfo {year}
  {2023})}\BibitemShut {NoStop}%
\bibitem [{\citenamefont {Kikuchi}\ \emph {et~al.}(2022)\citenamefont {Kikuchi}, \citenamefont {Matsukizono}, \citenamefont {Iwamatsu}, \citenamefont {Endo}, \citenamefont {Anan},\ and\ \citenamefont {Okumura}}]{original_smaf}%
  \BibitemOpen
  \bibfield  {author} {\bibinfo {author} {\bibfnamefont {H.}~\bibnamefont {Kikuchi}}, \bibinfo {author} {\bibfnamefont {H.}~\bibnamefont {Matsukizono}}, \bibinfo {author} {\bibfnamefont {K.}~\bibnamefont {Iwamatsu}}, \bibinfo {author} {\bibfnamefont {S.}~\bibnamefont {Endo}}, \bibinfo {author} {\bibfnamefont {S.}~\bibnamefont {Anan}}, \ and\ \bibinfo {author} {\bibfnamefont {Y.}~\bibnamefont {Okumura}},\ }\href {\doibase https://doi.org/10.1002/advs.202202048} {\bibfield  {journal} {\bibinfo  {journal} {Advanced Science}\ }\textbf {\bibinfo {volume} {9}},\ \bibinfo {pages} {2202048} (\bibinfo {year} {2022})}\BibitemShut {NoStop}%
\bibitem [{\citenamefont {Chen}\ \emph {et~al.}(2022)\citenamefont {Chen}, \citenamefont {Martinez}, \citenamefont {Nacke}, \citenamefont {Korblova}, \citenamefont {Manabe}, \citenamefont {Klasen-Memmer}, \citenamefont {Freychet}, \citenamefont {Zhernenkov}, \citenamefont {Glaser}, \citenamefont {Radzihovsky}, \citenamefont {Maclennan}, \citenamefont {Walba}, \citenamefont {Bremer}, \citenamefont {Giesselmann},\ and\ \citenamefont {Clark}}]{chen_smectic_a}%
  \BibitemOpen
  \bibfield  {author} {\bibinfo {author} {\bibfnamefont {X.}~\bibnamefont {Chen}}, \bibinfo {author} {\bibfnamefont {V.}~\bibnamefont {Martinez}}, \bibinfo {author} {\bibfnamefont {P.}~\bibnamefont {Nacke}}, \bibinfo {author} {\bibfnamefont {E.}~\bibnamefont {Korblova}}, \bibinfo {author} {\bibfnamefont {A.}~\bibnamefont {Manabe}}, \bibinfo {author} {\bibfnamefont {M.}~\bibnamefont {Klasen-Memmer}}, \bibinfo {author} {\bibfnamefont {G.}~\bibnamefont {Freychet}}, \bibinfo {author} {\bibfnamefont {M.}~\bibnamefont {Zhernenkov}}, \bibinfo {author} {\bibfnamefont {M.~A.}\ \bibnamefont {Glaser}}, \bibinfo {author} {\bibfnamefont {L.}~\bibnamefont {Radzihovsky}}, \bibinfo {author} {\bibfnamefont {J.~E.}\ \bibnamefont {Maclennan}}, \bibinfo {author} {\bibfnamefont {D.~M.}\ \bibnamefont {Walba}}, \bibinfo {author} {\bibfnamefont {M.}~\bibnamefont {Bremer}}, \bibinfo {author} {\bibfnamefont {F.}~\bibnamefont {Giesselmann}}, \ and\ \bibinfo {author} {\bibfnamefont {N.~A.}\ \bibnamefont {Clark}},\ }\href {\doibase
  10.1073/pnas.2210062119} {\bibfield  {journal} {\bibinfo  {journal} {Proceedings of the National Academy of Sciences}\ }\textbf {\bibinfo {volume} {119}},\ \bibinfo {pages} {e2210062119} (\bibinfo {year} {2022})}\BibitemShut {NoStop}%
\bibitem [{\citenamefont {Etxebarria}\ \emph {et~al.}(2022)\citenamefont {Etxebarria}, \citenamefont {Folcia},\ and\ \citenamefont {Ortega}}]{etxebarria_model}%
  \BibitemOpen
  \bibfield  {author} {\bibinfo {author} {\bibfnamefont {J.}~\bibnamefont {Etxebarria}}, \bibinfo {author} {\bibfnamefont {C.~L.}\ \bibnamefont {Folcia}}, \ and\ \bibinfo {author} {\bibfnamefont {J.}~\bibnamefont {Ortega}},\ }\href {\doibase 10.1080/02678292.2022.2055181} {\bibfield  {journal} {\bibinfo  {journal} {Liquid Crystals}\ }\textbf {\bibinfo {volume} {49}},\ \bibinfo {pages} {1719} (\bibinfo {year} {2022})}\BibitemShut {NoStop}%
\bibitem [{\citenamefont {Folcia}\ \emph {et~al.}(2022)\citenamefont {Folcia}, \citenamefont {Ortega}, \citenamefont {Vidal}, \citenamefont {Sierra},\ and\ \citenamefont {Etxebarria}}]{folcia_ferroelectric_2022}%
  \BibitemOpen
  \bibfield  {author} {\bibinfo {author} {\bibfnamefont {C.~L.}\ \bibnamefont {Folcia}}, \bibinfo {author} {\bibfnamefont {J.}~\bibnamefont {Ortega}}, \bibinfo {author} {\bibfnamefont {R.}~\bibnamefont {Vidal}}, \bibinfo {author} {\bibfnamefont {T.}~\bibnamefont {Sierra}}, \ and\ \bibinfo {author} {\bibfnamefont {J.}~\bibnamefont {Etxebarria}},\ }\href {\doibase 10.1080/02678292.2022.2056927} {\bibfield  {journal} {\bibinfo  {journal} {Liquid Crystals}\ }\textbf {\bibinfo {volume} {49}},\ \bibinfo {pages} {899} (\bibinfo {year} {2022})}\BibitemShut {NoStop}%
\bibitem [{\citenamefont {Gibb}\ \emph {et~al.}(2024)\citenamefont {Gibb}, \citenamefont {Hobbs}, \citenamefont {Nikolova}, \citenamefont {Raistrick}, \citenamefont {Gleeson},\ and\ \citenamefont {Mandle}}]{mandle_compound_1}%
  \BibitemOpen
  \bibfield  {author} {\bibinfo {author} {\bibfnamefont {C.}~\bibnamefont {Gibb}}, \bibinfo {author} {\bibfnamefont {J.}~\bibnamefont {Hobbs}}, \bibinfo {author} {\bibfnamefont {D.}~\bibnamefont {Nikolova}}, \bibinfo {author} {\bibfnamefont {T.}~\bibnamefont {Raistrick}}, \bibinfo {author} {\bibfnamefont {H.}~\bibnamefont {Gleeson}}, \ and\ \bibinfo {author} {\bibfnamefont {R.}~\bibnamefont {Mandle}},\ }\href@noop {} {\enquote {\bibinfo {title} {Spontaneous symmetry breaking in polar fluids},}\ } (\bibinfo {year} {2024}),\ \Eprint {http://arxiv.org/abs/2402.07305} {arXiv:2402.07305 [cond-mat.soft]} \BibitemShut {NoStop}%
\bibitem [{\citenamefont {Hobbs}\ \emph {et~al.}(2024)\citenamefont {Hobbs}, \citenamefont {Gibb},\ and\ \citenamefont {Mandle}}]{mandle_hobbs}%
  \BibitemOpen
  \bibfield  {author} {\bibinfo {author} {\bibfnamefont {J.}~\bibnamefont {Hobbs}}, \bibinfo {author} {\bibfnamefont {C.~J.}\ \bibnamefont {Gibb}}, \ and\ \bibinfo {author} {\bibfnamefont {R.~J.}\ \bibnamefont {Mandle}},\ }\href@noop {} {\enquote {\bibinfo {title} {Emergent anti-ferroelectric ordering and the coupling of liquid crystalline and polar order},}\ } (\bibinfo {year} {2024}),\ \Eprint {http://arxiv.org/abs/2404.12271} {arXiv:2404.12271 [cond-mat.soft]} \BibitemShut {NoStop}%
\bibitem [{\citenamefont {Matsukizono}\ \emph {et~al.}(2024)\citenamefont {Matsukizono}, \citenamefont {Sakamoto}, \citenamefont {Okumura},\ and\ \citenamefont {Kikuchi}}]{kikuchi_n_smaf}%
  \BibitemOpen
  \bibfield  {author} {\bibinfo {author} {\bibfnamefont {H.}~\bibnamefont {Matsukizono}}, \bibinfo {author} {\bibfnamefont {Y.}~\bibnamefont {Sakamoto}}, \bibinfo {author} {\bibfnamefont {Y.}~\bibnamefont {Okumura}}, \ and\ \bibinfo {author} {\bibfnamefont {H.}~\bibnamefont {Kikuchi}},\ }\href {\doibase 10.1021/acs.jpclett.3c03492} {\bibfield  {journal} {\bibinfo  {journal} {The Journal of Physical Chemistry Letters}\ }\textbf {\bibinfo {volume} {15}},\ \bibinfo {pages} {4212} (\bibinfo {year} {2024})}\BibitemShut {NoStop}%
\bibitem [{\citenamefont {Manabe}\ \emph {et~al.}(2021)\citenamefont {Manabe}, \citenamefont {Bremer},\ and\ \citenamefont {Kraska}}]{manabe_original}%
  \BibitemOpen
  \bibfield  {author} {\bibinfo {author} {\bibfnamefont {A.}~\bibnamefont {Manabe}}, \bibinfo {author} {\bibfnamefont {M.}~\bibnamefont {Bremer}}, \ and\ \bibinfo {author} {\bibfnamefont {M.}~\bibnamefont {Kraska}},\ }\href {\doibase 10.1080/02678292.2021.1921867} {\bibfield  {journal} {\bibinfo  {journal} {Liquid Crystals}\ }\textbf {\bibinfo {volume} {48}},\ \bibinfo {pages} {1079} (\bibinfo {year} {2021})}\BibitemShut {NoStop}%
\bibitem [{\citenamefont {Nishikawa}\ \emph {et~al.}(2023)\citenamefont {Nishikawa}, \citenamefont {Kuwayama}, \citenamefont {Nihonyanagi}, \citenamefont {Dhara},\ and\ \citenamefont {Araoka}}]{iso_smaf}%
  \BibitemOpen
  \bibfield  {author} {\bibinfo {author} {\bibfnamefont {H.}~\bibnamefont {Nishikawa}}, \bibinfo {author} {\bibfnamefont {M.}~\bibnamefont {Kuwayama}}, \bibinfo {author} {\bibfnamefont {A.}~\bibnamefont {Nihonyanagi}}, \bibinfo {author} {\bibfnamefont {B.}~\bibnamefont {Dhara}}, \ and\ \bibinfo {author} {\bibfnamefont {F.}~\bibnamefont {Araoka}},\ }\href {\doibase 10.1039/D3TC02212A} {\bibfield  {journal} {\bibinfo  {journal} {J. Mater. Chem. C}\ }\textbf {\bibinfo {volume} {11}},\ \bibinfo {pages} {12525} (\bibinfo {year} {2023})}\BibitemShut {NoStop}%
\bibitem [{\citenamefont {Song}\ \emph {et~al.}(2022)\citenamefont {Song}, \citenamefont {Deng}, \citenamefont {Wang}, \citenamefont {Li}, \citenamefont {Lei}, \citenamefont {Wan}, \citenamefont {Xia}, \citenamefont {Aya},\ and\ \citenamefont {Huang}}]{diocn}%
  \BibitemOpen
  \bibfield  {author} {\bibinfo {author} {\bibfnamefont {Y.}~\bibnamefont {Song}}, \bibinfo {author} {\bibfnamefont {M.}~\bibnamefont {Deng}}, \bibinfo {author} {\bibfnamefont {Z.}~\bibnamefont {Wang}}, \bibinfo {author} {\bibfnamefont {J.}~\bibnamefont {Li}}, \bibinfo {author} {\bibfnamefont {H.}~\bibnamefont {Lei}}, \bibinfo {author} {\bibfnamefont {Z.}~\bibnamefont {Wan}}, \bibinfo {author} {\bibfnamefont {R.}~\bibnamefont {Xia}}, \bibinfo {author} {\bibfnamefont {S.}~\bibnamefont {Aya}}, \ and\ \bibinfo {author} {\bibfnamefont {M.}~\bibnamefont {Huang}},\ }\href {\doibase 10.1021/acs.jpclett.2c02846} {\bibfield  {journal} {\bibinfo  {journal} {The Journal of Physical Chemistry Letters}\ }\textbf {\bibinfo {volume} {13}},\ \bibinfo {pages} {9983–9990} (\bibinfo {year} {2022})}\BibitemShut {NoStop}%
\bibitem [{\citenamefont {Madhusudana}(2021)}]{madhusudana}%
  \BibitemOpen
  \bibfield  {author} {\bibinfo {author} {\bibfnamefont {N.~V.}\ \bibnamefont {Madhusudana}},\ }\href {\doibase 10.1103/PhysRevE.104.014704} {\bibfield  {journal} {\bibinfo  {journal} {Physical Review E}\ }\textbf {\bibinfo {volume} {104}},\ \bibinfo {pages} {014704} (\bibinfo {year} {2021})}\BibitemShut {NoStop}%
\bibitem [{\citenamefont {Erkoreka}\ \emph {et~al.}(2024)\citenamefont {Erkoreka}, \citenamefont {Sebastián}, \citenamefont {Mertelj},\ and\ \citenamefont {Martinez-Perdiguero}}]{erkoreka_uuu}%
  \BibitemOpen
  \bibfield  {author} {\bibinfo {author} {\bibfnamefont {A.}~\bibnamefont {Erkoreka}}, \bibinfo {author} {\bibfnamefont {N.}~\bibnamefont {Sebastián}}, \bibinfo {author} {\bibfnamefont {A.}~\bibnamefont {Mertelj}}, \ and\ \bibinfo {author} {\bibfnamefont {J.}~\bibnamefont {Martinez-Perdiguero}},\ }\href {\doibase https://doi.org/10.1016/j.molliq.2024.125188} {\bibfield  {journal} {\bibinfo  {journal} {Journal of Molecular Liquids}\ }\textbf {\bibinfo {volume} {407}},\ \bibinfo {pages} {125188} (\bibinfo {year} {2024})}\BibitemShut {NoStop}%
\bibitem [{\citenamefont {Osipov}(2024{\natexlab{a}})}]{osipov}%
  \BibitemOpen
  \bibfield  {author} {\bibinfo {author} {\bibfnamefont {M.}~\bibnamefont {Osipov}},\ }\href {\doibase 10.1080/02678292.2024.2349667} {\bibfield  {journal} {\bibinfo  {journal} {Liquid Crystals}\ ,\ \bibinfo {pages} {1–7}} (\bibinfo {year} {2024}{\natexlab{a}})}\BibitemShut {NoStop}%
\bibitem [{\citenamefont {Osipov}(2024{\natexlab{b}})}]{osipov_review}%
  \BibitemOpen
  \bibfield  {author} {\bibinfo {author} {\bibfnamefont {M.}~\bibnamefont {Osipov}},\ }\href {\doibase 10.1080/21680396.2024.2360391} {\bibfield  {journal} {\bibinfo  {journal} {Liquid Crystals Reviews}\ }\textbf {\bibinfo {volume} {12}},\ \bibinfo {pages} {14–29} (\bibinfo {year} {2024}{\natexlab{b}})}\BibitemShut {NoStop}%
\bibitem [{\citenamefont {Chrzanowska}\ and\ \citenamefont {Longa}(2024)}]{dipoles_nf}%
  \BibitemOpen
  \bibfield  {author} {\bibinfo {author} {\bibfnamefont {A.}~\bibnamefont {Chrzanowska}}\ and\ \bibinfo {author} {\bibfnamefont {L.}~\bibnamefont {Longa}},\ }\href {\doibase 10.1080/02678292.2024.2350043} {\bibfield  {journal} {\bibinfo  {journal} {Liquid Crystals}\ ,\ \bibinfo {pages} {1–14}} (\bibinfo {year} {2024})}\BibitemShut {NoStop}%
\end{thebibliography}%

\end{document}